\documentclass{article}

\usepackage[english]{babel}

\usepackage[a4paper,top=2cm,bottom=2cm,left=3cm,right=3cm,marginparwidth=1.75cm]{geometry}

\usepackage{float}
\usepackage{amssymb}
\usepackage{siunitx}
\PassOptionsToPackage{hyphens}{url}\usepackage{hyperref}
\usepackage{cleveref}
\usepackage[utf8]{inputenc}
\usepackage[right]{lineno}
\usepackage{csquotes}
\usepackage{booktabs}
\usepackage{longtable}
\usepackage{adjustbox}
\usepackage{array}
\usepackage{url}
\usepackage{titlesec}
\usepackage{authblk}
\usepackage{xcolor} 

\usepackage[english]{babel}
\usepackage[style=numeric-comp,sorting=none,backend=biber,natbib=true,maxnames=2]{biblatex}
\DeclareNameAlias{sortname}{family-given}
\DeclareNameAlias{default}{family-given}

\renewbibmacro{in:}{}
\DeclareFieldFormat[article]{title}{\mkbibquote{#1}\addcomma}
\DeclareFieldFormat[book]{title}{\mkbibemph{#1}\addcomma}
\DeclareFieldFormat[bookinbook]{title}{\mkbibemph{#1}\addcomma}
\DeclareFieldFormat[inbook]{title}{\mkbibquote{#1}\addcomma}
\DeclareFieldFormat[incollection]{title}{\mkbibquote{#1}\addcomma}
\DeclareFieldFormat[inproceedings]{title}{\mkbibquote{#1}\addcomma}
\DeclareFieldFormat[manual]{title}{\mkbibemph{#1}\addcomma}
\DeclareFieldFormat[misc]{title}{\mkbibemph{#1}\addcomma}
\DeclareFieldFormat[thesis]{title}{\mkbibemph{#1}\addcomma}
\DeclareFieldFormat[unpublished]{title}{\mkbibquote{#1}\addcomma}
\DeclareFieldFormat[patent]{title}{\mkbibemph{#1}\addcomma}
\DeclareFieldFormat[report]{title}{\mkbibemph{#1}\addcomma}
\DeclareFieldFormat[online]{title}{\mkbibquote{#1}\addcomma}
\DeclareFieldFormat[software]{title}{\mkbibemph{#1}\addcomma}
\DeclareFieldFormat[booklet]{title}{\mkbibemph{#1}\addcomma}
\DeclareFieldFormat[periodical]{title}{\mkbibemph{#1}\addcomma}
\DeclareFieldFormat[standard]{title}{\mkbibemph{#1}\addcomma}

\DeclareFieldFormat[article]{journaltitle}{\iffieldundef{shortjournal}{\mkbibemph{#1}\addcomma}{\mkbibemph{\printfield{shortjournal}}\addcomma}}
\DeclareFieldFormat{volume}{\bibstring{volume}~#1}
\DeclareFieldFormat{number}{\bibstring{number}~#1}

\DefineBibliographyStrings{english}{
  volume = {Vol.},
  number = {No.}
}

\renewbibmacro*{volume+number+eid}{%
  \printfield{volume}%
  \setunit*{\addspace}%
  \printfield{number}%
  \setunit{\addcomma\space}%
  \printfield{eid}}

\renewbibmacro*{journal+issuetitle}{%
  \usebibmacro{journal}%
  \setunit*{\addcomma\space}%
  \usebibmacro{volume+number+eid}%
  \setunit{\addcomma\space}%
  \usebibmacro{issue+date}}

\renewbibmacro*{publisher+location+date}{%
  \printlist{publisher}%
  \iflistundef{location}
    {\setunit*{\addcomma\space}}
    {\setunit*{\addcolon\space}}%
  \printlist{location}%
  \setunit*{\addcomma\space}%
  \usebibmacro{date}}

\DefineBibliographyStrings{english}{
  andothers = {\textit{et al.},}
}

\DeclareFieldFormat[article]{volume}{\bibstring{jourvol}\addnbspace #1}
\DeclareFieldFormat[article]{number}{\bibstring{number}\addnbspace #1}
\DeclareFieldFormat[article]{volume}{Vol. #1}
\DeclareFieldFormat[article]{number}{No. #1}

\DeclareFieldFormat{url}{\bibstring{available at}\addcolon\space\url{#1}}
\DeclareFieldFormat{urldate}{\mkbibparens{accessed \addspace#1}}

\DeclareFieldFormat{urldate}{%
  \mkbibparens{accessed\space%
    \thefield{urlday}\addspace%
    \mkbibmonth{\thefield{urlmonth}}\addspace%
    \thefield{urlyear}}}

\crefformat{figure}{#2Figure~#1#3}
\Crefformat{figure}{#2Figure~#1#3}
\crefformat{table}{#2Table~#1#3}
\Crefformat{table}{#2Table~#1#3}
\crefformat{section}{#2Section~#1#3}
\Crefformat{section}{#2Section~#1#3}

\author[1]{Seongeun Yun}
\author[1*]{Won Bo Lee}

\affil[1]{Seoul National University, Chemical and Biological Engineering, Seoul, Korea}
\affil[]{*Corresponding author E-mail : wblee@snu.ac.kr}

\title{Hierarchical Framework for Retrosynthesis Prediction 
with Enhanced Reaction Center Localization
}

\begin{document}
\maketitle

\begin{abstract}
Retrosynthesis is essential for designing synthetic pathways for complex molecules and can be revolutionized by AI to automate and accelerate chemical synthesis planning for drug discovery and materials science. Here, we propose a hierarchical framework for retrosynthesis prediction that systematically integrates reaction center identification, action prediction, and termination decision into a unified pipeline. Leveraging a molecular encoder pretrained with contrastive learning, the model captures both atom and bond level representations, enabling accurate identification of reaction centers and prediction of chemical actions. The framework addresses the scarcity of multiple reaction center data through augmentation strategies, enhancing the model’s ability to generalize to diverse reaction scenarios. The proposed approach achieves competitive performance across benchmark datasets, with notably high topk accuracy and exceptional reaction center identification capabilities, demonstrating its robustness in handling complex transformations. These advancements position the framework as a promising tool for future applications in material design and drug discovery.

\end{abstract}

\section{Introduction}
\label{sec:introduction}

Computer-aided synthesis planning (CASP)\cite{b1, b2, b3} has become an essential tool in domains such as drug discovery, material design, and synthetic chemistry, enabling the efficient identification of synthetic pathways for molecular compounds and accelerating material development. Early CASP systems predominantly relied on expert-developed rules to guide synthesis tasks. For instance, LHASA\cite{b3} utilizes reaction templates to construct candidate synthesis trees; however, these templates often fail to account for the broader molecular context, leading to inaccuracies in forward synthesis predictions. Similarly, systems such as SOPHIA\cite{b4} and Eros\cite{b5} employ curated reactivity rules to identify reactive functional groups, while Chematica’s Syntaurus\cite{b6} incorporates predefined lists of incompatible functional groups associated with retrosynthetic templates. Although effective in constrained scenarios, these rule-based approaches are limited by their dependence on manually encoded rules, restricting their ability to predict novel reactions or operate beyond predefined parameters. To address these limitations, there has been a growing interest in the application of machine learning techniques, offering the potential to enhance CASP by enabling flexible, context-aware synthesis planning and broadening the scope of its predictive capabilities.\cite{b6}
\\

\noindent The demonstrated ability of artificial intelligence (AI) to surpass human performance in various domains, alongside the increasing accessibility of extensive reaction databases, has generated significant interest in AI-driven retrosynthesis prediction.\cite{b6} In response, a variety of approaches have been developed, including template-based, template-free, and semi-template-based methodologies.

Template-based methods\cite{b8, b9, b10, b11, b12} utilize reaction templates that encode fundamental rules for transforming products into reactants. These templates, often generated automatically, serve as a basis for classifying product molecules and identifying the corresponding reactants. Molecular representations in this approach can take various forms, such as graphs, textual notations, or molecular fingerprints. For example, Segler and Waller used Morgan Fingerprints to classify reaction templates\cite{b8}, employing Deep Neural Network (DNN) for input representation. The Graph Logic Network (GLN)\cite{b9} enhances template-based methods by integrating reaction templates with a conditional graphical model, allowing for the dynamic learning of template application rules while ensuring chemical feasibility and strategic alignment. LocalRetro\cite{b10}, a state-of-the-art template-based model, improves accuracy by prioritizing atom- and bond-level local templates and applying them to predicted reaction centers. One notable advantage of template-based methods is their interpretability. The use of templates provides a clear understanding of the model’s predictions and the underlying reaction mechanisms. However, these methods face significant limitations due to the finite number of templates available and the associated computational constraints. This restricts their ability to cover the full spectrum of possible reactions and limits flexibility. Additionally, predictions outside the scope of established templates are not feasible, which can result in reduced accuracy compared to other approaches.

Template-free methods\cite{b13, b14, b15, b16, b17, b18, b19, b20} directly predict reactants from products without relying on predefined reaction templates, utilizing models that learn complex mappings between molecular structures. This approach effectively circumvents the limitations imposed by template databases. A foundational method in this category is the Seq2Seq model with Long Short-Term Memory (LSTM)\cite{b13}, which translates product SMILES strings into reactant SMILES\cite{b14}. Despite its utility, it faced issues with generating invalid chemical structures. SCROP\cite{b15} improved upon Transformer\cite{b16} models by introducing a grammar correction module, addressing syntax errors in predicted SMILES and enhancing the validity of the outputs. Retroformer\cite{b17} advanced this paradigm further by jointly encoding molecular sequences and graphs via a local attention mechanism, enabling interpretable and controllable generative processes without reliance on cheminformatics tools. The primary advantage of template-free methods lies in their flexibility, allowing them to generalize to a broader range of reactions, including novel and previously unseen scenarios. However, a significant limitation of these methods is their lack of interpretability, making it challenging to understand the reasoning behind their predictions. Additionally, the absence of built-in validity checks raises concerns about the reliability and accuracy of the predicted reactants, particularly for complex molecular structures.

Semi-template methods\cite{b21, b22, b23, b24, b25, b26, b27} address the limitations of template-based and template-free approaches by incorporating chemists’ intuition and decomposing retrosynthesis into manageable steps. These methods leverage the principle that most reactions involve modifications to a localized region of the molecule, predicting the final reactant through a sequential process of reaction center identification and synthon completion.\cite{b21} Advanced techniques such as graph generators, transformers, and subgraph selection models are employed to efficiently reconstruct reactants from the product molecule. In G2G\cite{b22}, the reaction center is restricted to newly formed bonds, limiting its ability to handle diverse reaction data comprehensively. Additionally, its reliance on a graph generative approach for synthon completion has shown lower accuracy compared to other semi-template methods. RetroPrime\cite{b23} addresses these limitations by expanding reaction center identification to consider all atoms within the molecule as potential candidates. Using a transformer-based architecture, RetroPrime predicts synthons and converts them into final reactants, effectively managing complex molecular transformations. These advancements demonstrate the potential of semi-template methods to balance flexibility and interpretability in retrosynthesis prediction. For reactions where reactants contain atoms not existing in the product, certain methods treat these as atom groups and incorporate them using a predefined atom group library extracted from the dataset. This approach\cite{b24} ensures the systematic and efficient inclusion of such groups into the retrosynthetic process. RetroExplainer\cite{b25} adopts a mechanistic framework inspired by the Sn2 reaction mechanism, employing specialized modules to predict bond changes, leaving group attachments, and hydrogen atom count adjustments, enabling one-step retrosynthetic predictions from product to reactant. While semi-template methods offer diverse strategies to address retrosynthesis challenges, many focus exclusively on bond reaction centers or consider both atom and bond reaction centers without effectively integrating their predictions. The separate modeling of atom centers and bond centers introduces challenges in systematic integration, often resulting in suboptimal performance. This lack of a cohesive framework limits the ability to capture the intricate relationships between atom-level and bond-level changes, highlighting a critical area for improvement in existing approaches.
\\

\begin{figure}[H]
 \centering
 \makebox[\textwidth][c]{\includegraphics[width=1\textwidth]{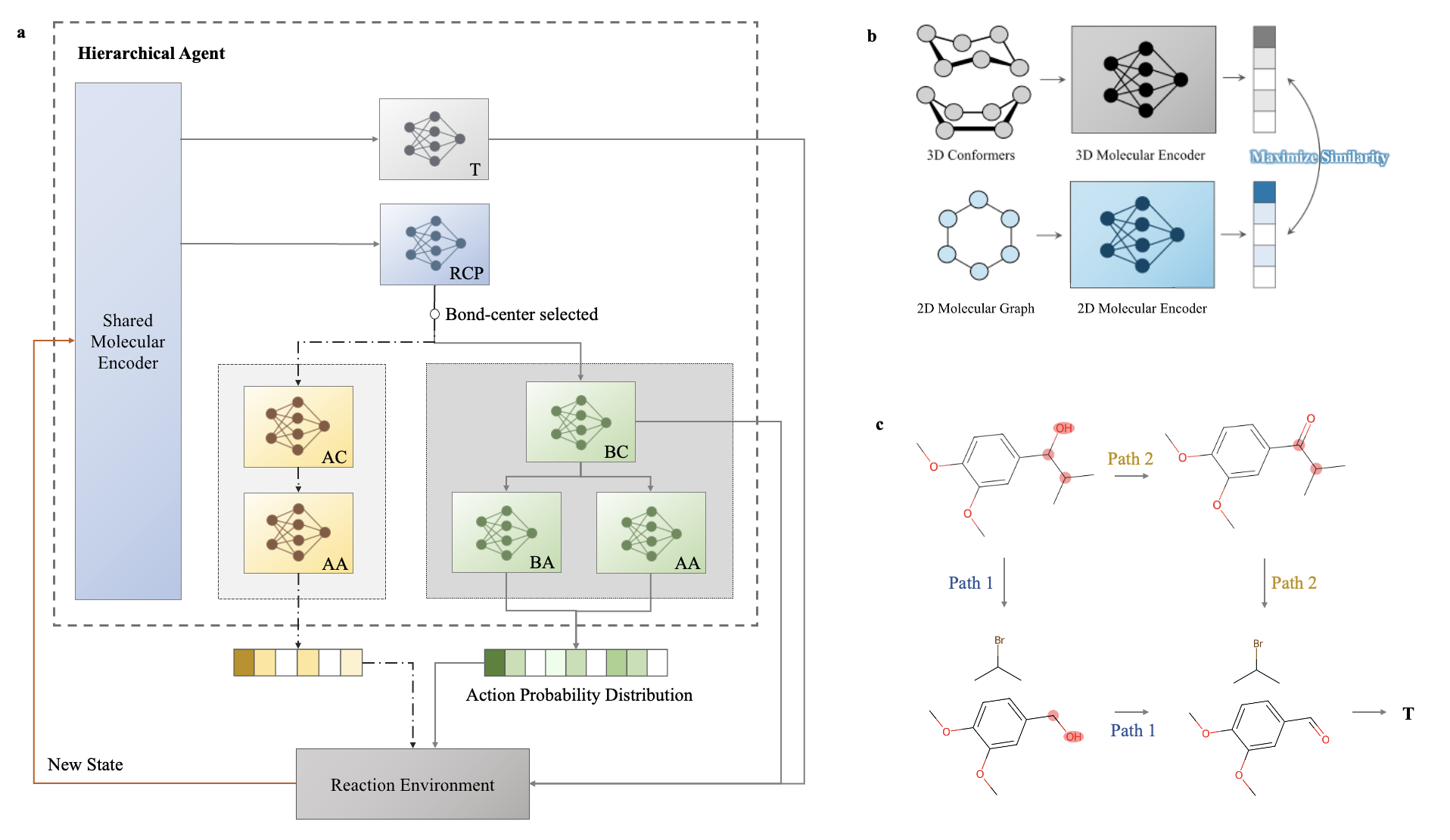}}%
 \caption{\textbf{Overview of HierRetro}, \textbf{a} The architecture of HierRetro for retrosynthesis prediction. The shared molecular encoder generates atom-level and pair-level representations, which are utilized across all modules. The Reaction Center Type Prediction (RCP) module determines whether the reaction center is an atom or a bond, directing the flow to the Atom Center (AC) or Bond Center (BC) modules for localization. Action Prediction modules (AA and BA) then determine the chemical modifications at the identified reaction center. Finally, the Termination (T) module evaluates whether the process should conclude or proceed to the next reaction step. The action probability distribution generated at each step guides updates in the reaction environment, enabling iterative reasoning. \textbf{b} The molecular encoder was pre-trained using contrastive learning (CL) by maximizing the similarity between outputs from a 2D network, utilizing 2D molecular graphs, and a 3D network, incorporating conformer representations.  This approach allows the encoder to implicitly learn 3D information for enhanced downstream performance. \textbf{c} Retrosynthetic pathways for a multi-reaction center molecule. Two distinct reaction centers are identified, leading to two potential retrosynthetic pathways (Path 1 and Path 2). The final product is achieved via termination (T) following the completion of each respective pathway.}
 \label{fig-1}
\end{figure}

\noindent In this work, we propose a hierarchical framework, HierRetro, with two key features:

First, we present a hierarchical model for retrosynthesis prediction designed to systematically address the inherent complexity of molecular transformations, as illustrated in Fig.1. The model begins by classifying the reaction center type, distinguishing between atom and bond centers to enable structured and efficient decision-making when both types are considered. Following this, the precise location of the reaction center is identified based on the classified type, and the necessary edit is predicted using the features associated with the identified center. Finally, the model determines whether to terminate the process or proceed, enabling the effective handling of multiple reaction centers and ensuring adaptability to complex molecular transformations. By decomposing retrosynthesis into distinct and interpretable modules, this approach enhances prediction accuracy while maintaining flexibility in addressing a diverse range of molecular reactions.

Additionally, HierRetro incorporates contrastive learning\cite{b28, b29, b30} to pretrain a molecular encoder, enabling it to implicitly capture the 3-dimensional structural information of input molecules while using solely 2-dimensional molecular graphs, as illustrated in Fig. 2. Contrastive learning, an unsupervised learning technique, works by maximizing the similarity between hidden features derived from various augmented versions of the same input. In the 3D-InfoMax\cite{b29} approach, the model employs a 2D network, which uses molecular connectivity information, and a 3D network, which leverages the molecule’s 3D coordinates. By maximizing the agreement between the hidden features generated by these two networks, the 2D network is trained to encode 3D structural information implicitly. The USPTO dataset, widely used as a benchmark for reaction prediction tasks, provides molecular data exclusively in SMILES format, lacking explicit 3D structural information. Although 3D coordinates can be generated using external methods, each has notable limitations. Quantum chemistry techniques such as density functional theory (DFT)\cite{b31, b32} provide high accuracy but are computationally expensive and unsuitable for large-scale datasets. Faster alternatives like the MMFF\cite{b33} method in RDKit\cite{b34} reduce computational costs but sacrifice accuracy. Moreover, directly incorporating 3D coordinates into the model increases the number of parameters, which can lead to reduced predictive performance.25 In some instances, the addition of explicit 3D information has even been observed to decrease model accuracy. By leveraging contrastive learning, our model circumvents these challenges, offering an efficient and effective means of capturing 3D structural information without the computational and accuracy trade-offs associated with explicit 3D data.

\section{Method}
\label{sec:method}

\subsection{Data Preparation}
In this study, we utilized the USPTO-50k benchmark dataset, a widely recognized resource containing atom-mapped reactions categorized into 10 distinct reaction types\cite{b35}. The dataset was divided into training, validation, and test sets, following the splits proposed by Coley et al.\cite{b12} To mitigate potential information leakage caused by the atom-mapping algorithm\cite{b36, b37}, the mapping numbers were randomized to reassign reaction center indices, and all SMILES strings were canonicalized to ensure consistency across the dataset.

Reactions were classified into two groups: those involving bond changes and those without. For each reaction, graph edits were manually extracted from the training dataset to capture the structural modifications accurately. Atom reaction center actions include possible edits such as changes in hydrogen count, chiral modifications, and atom group attachments. For bond reaction centers, actions were defined in two stages: the first stage involves bond-level modifications, such as bond type changes or bond deletions, while the second stage accounts for atom-level edits applied to the pair of atoms connected by the bond, capturing the full range of possible transformations. This systematic approach enables the extraction of action sets that comprehensively cover 99.9\% of the reactions in the validation and test sets, ensuring that the extracted actions effectively represent the diversity and complexity of reaction mechanisms within the dataset.

\subsection{Pretraining Molecular Encoder}
In this study, we adopt the state-of-the-art Uni-Mol+\cite{b38, b39} encoder architecture as the foundation for our molecular encoder. Uni-Mol+ utilizes a Transformer-based design to process atom-level and pair-level representations, capturing both local and global molecular features. Atom representations encode individual atomic properties, while pair representations combine bond features with 3D spatial and 2D topological information. Since the USPTO dataset lacks explicit 3D structures, pair representations in our framework only rely on bond features and 2D graph topology. The encoder updates both atom and pair representations using Transformer blocks, enabling efficient integration of local interactions and long-range dependencies. Compared to GNNs, Uni-Mol+ excels in scalability, accuracy, and flexibility, making it a robust backbone for retrosynthesis prediction.

Inspired by the 3D-Infomax\cite{b29} approach, we employ contrastive learning (CL) during pretraining to enhance the encoder’s ability to capture 3D structural insights without requiring explicit 3D data. This method leverages a dual-network setup: a 2D network utilizing molecular connectivity and a 3D network incorporating conformer information. Pretraining is performed using the GEOM-Drug dataset\cite{b40}, which contains 304,466 molecules annotated with diverse quantum mechanical (QM) properties and 3D conformers. During training, a random conformer structure is selected for each molecule at every epoch, ensuring robust learning across various molecular conformations. This pretraining strategy enables the encoder to implicitly encode 3D structural information while maintaining adaptability to datasets like USPTO, where only 2D representations are available. By integrating Uni-Mol+ architecture with contrastive learning, our encoder establishes a strong foundation for accurate and efficient retrosynthesis prediction.

\subsection{Multitasking Agent}
The multitasking agent in our framework is designed to address multiple interconnected tasks that are critical for retrosynthesis prediction. To obtain updated atom and pair representations at each step t, the molecular encoder incorporates information from the current state as well as the previous state. The updates for the atom representation $h_a^t$ and the pair representation $h_b^t$ are computed as follows:
\begin{eqnarray}
h_a^t=W_a^c h_a^t+W_a^p h_a^{t-1}\\
h_b^t=W_b^c h_b^t+W_b^p h_b^{t-1}
\end{eqnarray}
where $W^c$ is the weight matrix applied to the current representations and $W^p$ is the weight matrix applied to the previous representations. 

Utilizing these atom and pair representations, the agent performs three key tasks: reaction center identification, action prediction, and termination decision. Each task contributes to the retrosynthesis process with distinct functionalities and corresponding loss functions, as described below.

\subsubsection{Reaction Center Identification}
The reaction center identification process consists of three modules: Reaction Center Type Prediction (RCP), Atom Center Prediction (AC), and Bond Center Prediction (BC). These modules work together to predict the location of the reaction center within a molecule.

\textbf{RCP Module}
The RCP module determines whether the reaction center is an atom or a bond by utilizing the graph-level super node features $h_g^t$. the predicted probability $p_{RCP}^t$ for the reaction center being a bond is calculated as:
\begin{equation}
p_{RCP}^t=\sigma (W_{RCP} h_g^t )
\end{equation}

\textbf{AC Module}
The AC module predicts the atom reaction center among all atoms in the molecule using the atom representation $h_a^t$. The predicted probability distribution $p_{AC}^t$ over all atoms is calculated as:
\begin{equation}
p_{AC}^t=softmax(W_{AC} h_a^t )
\end{equation}

\textbf{BC Module}
The BC module predicts the bond reaction center among all atom pairs in the molecule using the flattened bond pair representations $h_b^t$. The predicted probability distribution $p_{BC}^t$ over all atom pairs is calculated as:
\begin{equation}
p_{BC}^t=softmax(W_{BC} h_b^t )
\end{equation}

The reasoning step incorporates the outputs of all three modules to calculate the overall reaction center probability $p_{RC}^t$. This combines the atom and bond center predictions weighted by the RCP probability:
\begin{equation}
p_{RC}^t=concat[(1-p_{RCP}^t )\centerdot p_{AC}^t, p_{RCP}^t \centerdot p_{BC}^t ]
\end{equation}

The loss for reaction center identification reflects the alignment between the predicted reaction center probability distribution $p_{RC}^t$ and the ground truth distribution, accounting for multiple reaction centers in the molecule. It is computed as:
\begin{equation}
\mathcal{L_{RC}}=-\sum_t \sum_{k \in RC^t} y^t_{RC,k} log p^t_{RC,k}
\end{equation}

where $y^t_{RC,k}$ is the ground truth probability for reaction center $k$ and $p^t_{RC,k}$ is the predicted probability for reaction center $k$. This approach ensures that the model systematically identifies the reaction center while incorporating the contributions of atom- and bond-level predictions.

\subsubsection{Action Prediction}
The action prediction process determines the specific chemical modifications required at the identified reaction center and consists of two modules: Atom Action Prediction (AA) and Bond Action Prediction (BA). Each module utilizes the reaction center information determined in the previous task to compute the appropriate action.

\textbf{AA Module}
The AA predicts the probability distribution of atom actions occurring at a given atom index. It uses the atom feature of the identified atom, $h^t_{a,i}$, which is extracted from the atom representation corresponding to the atom index, $i$. The predicted probability distribution $p^t_{AA,i}$  over possible atom actions is calculated as:
\begin{equation}
p^t_{AA,i}=softmax(W_{AA} h^t_{a,i} )
\end{equation}

\textbf{BA Module}
The BA module predicts the probability distribution of atom actions occurring at a given bond index. It uses the pair feature of the identified bond, $h_{b,ij}^t$, which is extracted from the pair representation corresponding to the bond center index, $ij$. The predicted probability distribution $p_{BA,ij}^t$ over possible bond actions is calculated as:
\begin{equation}
p_{BA,ij}^t=softmax(W_{BA} h_{b,ij}^t  )
\end{equation}

To fully determine the bond action, the model also considers atom-level actions on both ends of the identified bond. These actions are represented by $p_{AA,i}^t$ and $p_{AA,j}^t$,  the probabilities of atom actions for the two atoms involved in the bond.

The losses for the action prediction modules are calculated using cross-entropy loss:
Atom action loss is calculated with cross entropy loss as :
\begin{equation}
\mathcal{L}_{AA} = -\sum_t \log(p_{AA,i_t}^t)
\end{equation}

Bond action loss is calculated with cross entropy loss as :
\begin{equation}
\mathcal{L}_{BA} = -\sum_t \log(p_{BA,ij_t}^t) + \log(p_{AA,i_t}^t) + \log(p_{AA,j_t}^t)
\end{equation}

where $i_t$ and $ij_t$ are atom index and bond indices identified in the previous task at step t.

\subsubsection{Termination Determination}
The termination decision task predicts whether the retrosynthesis process should conclude or proceed to the next reaction step. This task is framed as a binary classification problem, utilizing the graph-level super node features, $h_g^t$. The termination probability $p_T^t$ is calculated as :
\begin{equation}
p_T^t=\sigma (W_T h_g^t)
\end{equation}

The loss for the termination decision task is calculated using binary cross-entropy. It penalizes the model based on the difference between the predicted termination probability $p_T^t$ and the ground truth $y_T^t$:
\begin{equation}
\mathcal{L}_T = -\sum_t \bigg( y_T^t \log(p_T^t) + (1-y_T^t) \log(1-p_T^t) \bigg)
\end{equation}

To integrate the four losses ($\mathcal{L_{RC}}$, $\mathcal{L_{AA}}$, $\mathcal{L_{BA}}$, and $\mathcal{L_{T}}$), we adopt a dynamic adaptive multi-task learning (DAMT)\cite{b25} strategy. Tasks in retrosynthesis prediction vary in complexity and loss magnitude, which can lead to conflicts in parameter updates and imbalance during training. By dynamically adjusting the loss weights based on task complexity and normalizing their magnitudes, this approach ensures that each task contributes appropriately to the optimization process. This strategy prevents dominant tasks from overshadowing others, enabling balanced learning and improving both convergence and generalization.

\subsection{Model Details}
HierRetro was trained using the AdamW\cite{b41} optimizer, which effectively balances learning and weight decay for stable optimization. A WarmUpWrapper combined with ReduceLROnPlateau was employed as the learning rate scheduler. This approach starts with a gradual warm-up phase to reach a peak learning rate of $2 \centerdot 10^{-6}$, followed by adaptive adjustments when validation performance plateaus, with a minimum learning rate set to $1 \centerdot 10^{-7}$. The architecture consists of 6 encoder blocks with a atom hidden dimension of 256 and pair hidden dimension of 128. Training was performed on an NVIDIA A100 GPU, with the process completing in approximately 42 hours.

\section{Results}
\label{sec:main_section}

\subsection{Performance Evaluation}
Top-k exact match accuracy was employed as the primary evaluation metric to compare the performance of HierRetro against existing approaches. To calculate top-k accuracy, a beam search was employed at each step and for each decision-making process within the step. The model’s performance was evaluated in two cases: reaction type unknown and reaction type known, with accuracy analyzed for k = 1, 3, 5, and 10. The comparative analysis included template-based methods (Retrosim\cite{b12}, Neuralsym\cite{b8}, GLN\cite{b9}, LocalRetro\cite{b10}), template-free methods (SCROP\cite{b15}, Retroformer\cite{b17}, GTA\cite{b18}, Graph2Smiles\cite{b20}, Dual\cite{b19}), and semi-template methods (G2G\cite{b22}, RetroXpert\cite{b21}, RetroPrime\cite{b23}, MEGAN\cite{b26}, GraphRetro\cite{b37}, Graph2Edits\cite{b24}). This comprehensive evaluation highlights the model’s competitive ability to predict accurate reactants, and the results are summarized in \textbf{Table 1}.

In the reaction type unknown case, HierRetro achieved a top-3 accuracy of 78.3\%, outperforming all other models. The model maintained competitive performance with a top-5 accuracy of 83.1\%, slightly behind Graph2Edits’ 83.4\%, and recorded a top-10 accuracy of 89.1\%. For the reaction type known case, the model achieved a top-3 accuracy of 89.4\%, 1.9\% higher than Graph2Edits (87.5\%), and a top-5 accuracy of 93.3\%, surpassing Graph2Edits and LocalRetro by 1.8\% and 0.9\%, respectively. The model recorded the highest top-10 accuracy of 94.9\% among semi-template methods. The superior performance of the model stems from its systematic reasoning process, which ensures accurate predictions. The hierarchical structure, which involves determining the reaction center type, locating the reaction center based on this decision, and subsequently integrating the results into the action prediction process, effectively guides the model towards identifying the correct solution. This stepwise reasoning framework ensures systematic and accurate retrosynthesis predictions. Furthermore, the enhanced performance with reaction type information highlights its ability to effectively leverage the correlation between reaction types and retrosynthesis tasks. The pre-trained encoder further contributes to this success by providing robust feature representations.

\begin{table}[ht]
    \centering
    \includegraphics[width=0.8\textwidth]{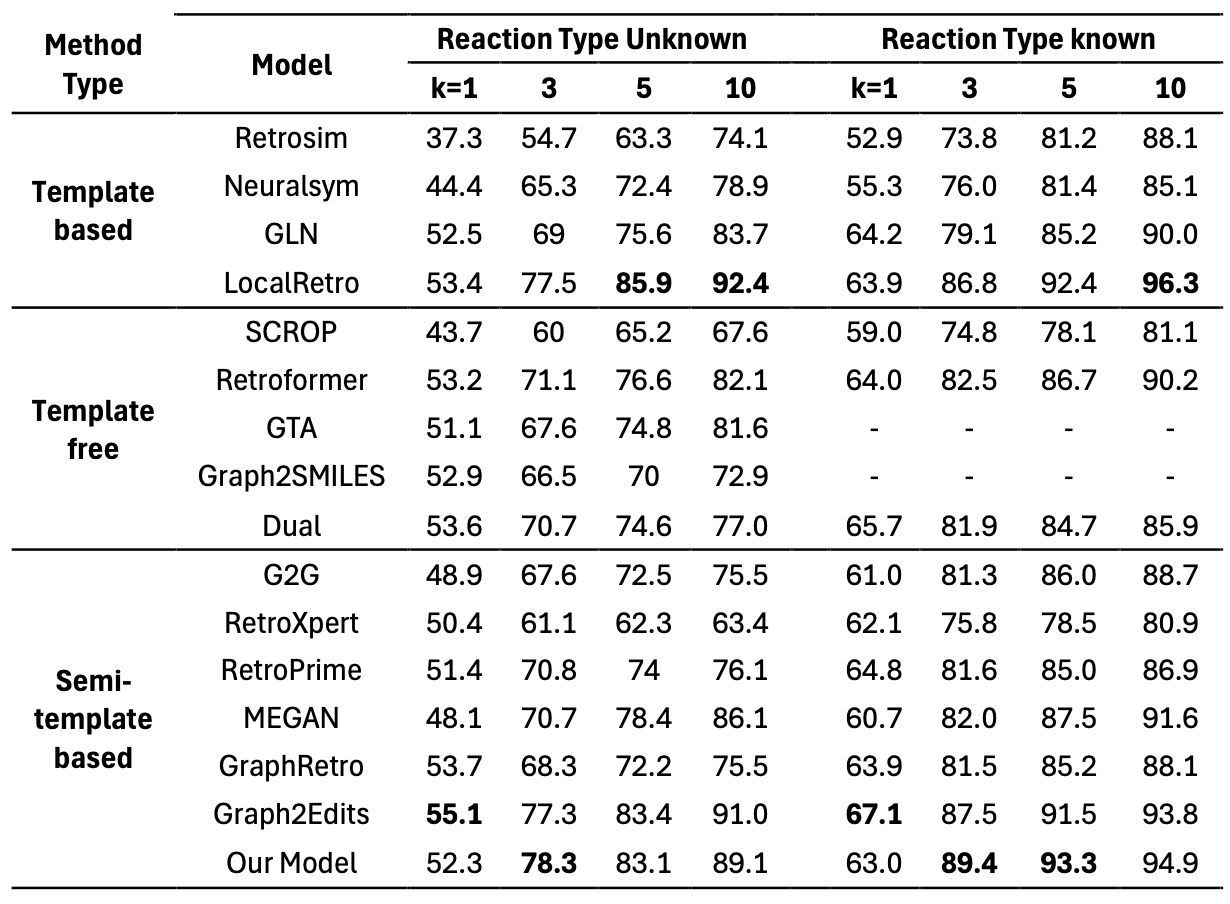} 
    \caption{Top-k exact accuracy on USPTO-50K benchmarks.}
    \label{tab-1}
\end{table}

\noindent Given the nature of reaction prediction tasks, we evaluated the model’s performance using round-trip\cite{b42} accuracy as a complementary metric to top-k exact match accuracy. Round-trip accuracy utilizes a forward reaction prediction model, such as Molecular Transformer\cite{b43}, to regenerate the product from the predicted reactants. The regenerated product is compared with the original product to determine whether the predicted reactants can plausibly reproduce it, even if they do not match the ground truth exactly. This metric addresses the strict matching criteria of exact match accuracy, which may incorrectly penalize valid predictions. For instance, if the model predicts Cl as a leaving group instead of the ground truth Br in an Sn2 reaction, the prediction is still chemically valid since Cl can also produce the same product. Round-trip accuracy resolves such cases by recognizing plausible reactants that lead to the correct product, offering a more comprehensive assessment of model performance. The evaluation of round-trip accuracy was conducted in comparison with existing methods, including LocalRetro\cite{b10}, MEGAN\cite{b26}, GraphRetro\cite{b37}, and Graph2Edits\cite{b24}. The results, summarized in \textbf{Table 2}, highlight the model’s capability to generate chemically valid reactants that align with the original product.

	As presented in \textbf{Table 2}, our model demonstrated competitive performance across various k-values in the round-trip accuracy evaluation. The model achieved 94.7\% accuracy for k=3 and 96.2\% for k=5, outperforming all models except LocalRetro. At k=10, HierRetro achieved the highest accuracy of 97.9\%, surpassing Graph2Edits and highlighting its robustness in generating valid reactant sets across broader candidate spaces. These results underscore the model’s ability to effectively generate plausible and realistic reactant sets for retrosynthesis tasks.

\begin{table}[ht]
 \centering
 \includegraphics[width=0.8\textwidth]{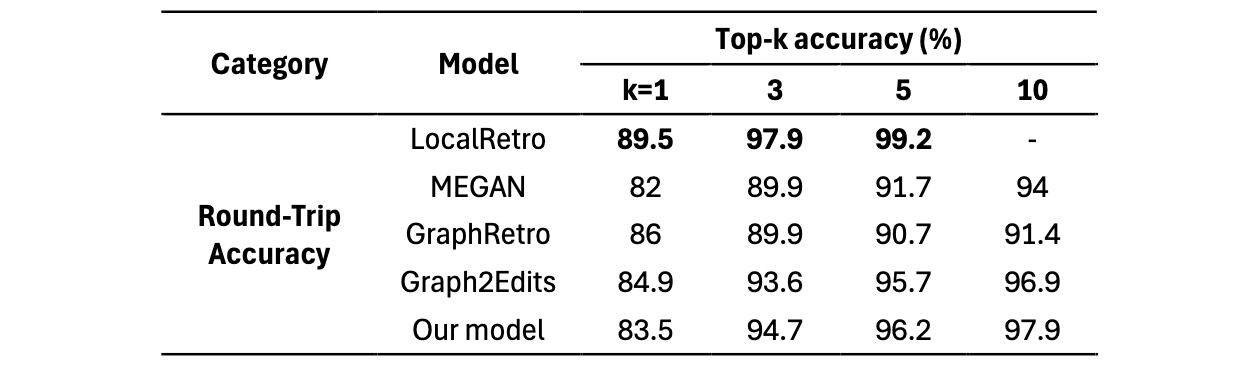} 
 \caption{Top-k round-trip accuracy on USPTO-50K benchmarks}
 \label{tab-2}
\end{table}

\subsection{Ablation Study}
An ablation study was conducted to assess the impact of incorporating a contrastive learning (CL)-based pre-trained encoder on the model’s performance under the reaction type unknown condition. With the pre-trained encoder, the model achieved a top-1 accuracy of 52.3\%, compared to 49.8\% without pre-training, marking a 2.5\% improvement. This enhancement is attributed to the pre-trained encoder’s ability to capture general patterns from larger datasets, providing a stronger foundation for learning on the smaller and less diverse USPTO dataset. Furthermore, the implicit integration of 3D structural information, challenging to infer from SMILES alone, improved performance without adding parameters or embeddings, minimizing risks of overfitting while enhancing model robustness.

	The role of the reaction center type prediction module was also evaluated. Without this module, the reaction center was selected based on the higher probability between atom and bond center predictions. Adding this module increased reaction center identification accuracy under the reaction type unknown condition to 72.5\%, compared to 67.5\% without it, an improvement of approximately 5\%. This difference arises from the encoder’s information exchange between atom and bond representations, which causes atoms and bonds near the reaction center to both receive higher scores, making it challenging to distinguish between the two. By explicitly predicting the reaction center type and subsequently identifying its location, the model achieved higher accuracy and improved learning efficiency, demonstrating the effectiveness of this modular approach.

\subsection{Individual Module Performance}
The evaluation of the reaction center identification and synthon completion modules reveals the strong performance of our model in both reaction type unknown and reaction type known scenarios. The evaluation of both modules was conducted in comparison with GraphRetro\cite{b37}, RetroPrime\cite{b23}, G2Retro\cite{b27}, and the results are presented in \textbf{Table 3}.

	In the reaction center identification module without reaction type, our model achieved top-1, top-3, and top-5 accuracies of 72.5\%, 92.2\%, and 95.1\%, outperforming all compared semi-template models. When reaction type information was provided, the model demonstrated further improvements, achieving top-1, top-3, and top-5 accuracies of 85.0\%, 97.3\%, and 98.1\%, the highest among all compared methods. These outstanding results demonstrate that the hierarchical framework introduced in the model enables more accurate identification of reaction centers. Additionally, the significant increase in accuracy when reaction type information is provided highlights the model’s ability to effectively utilize reaction type data. This performance further underscores the encoder’s strength in exchanging atom and bond information and the impact of pretraining on enhancing feature representations.
    
	For synthon completion, the model achieved a top-3 accuracy of 93.1\% in the reaction type unknown scenario, outperforming other semi-template models. While the top-1 and top-5 accuracies were slightly lower than those of the best-scored model, they remained competitive. In the reaction type known scenario, the model achieved top-2 and top-3 accuracies of 90.5\% and 95.2\%, respectively, surpassing all other methods. The precise identification of reaction centers significantly contributed to accurate action prediction, enabling effective synthon completion. These results underscore the critical role of reliable reaction center identification in ensuring consistent and accurate retrosynthesis predictions.
    
	Overall, the results demonstrate that HierRetro consistently delivers high performance across both reaction center identification and synthon completion tasks. The performance improvements observed when reaction type information is provided further highlight the model’s capacity to leverage contextual information effectively, resulting in state-of-the-art performance in reaction center identification.

\begin{table}[ht]
 \centering
 \includegraphics[width=0.8\textwidth]{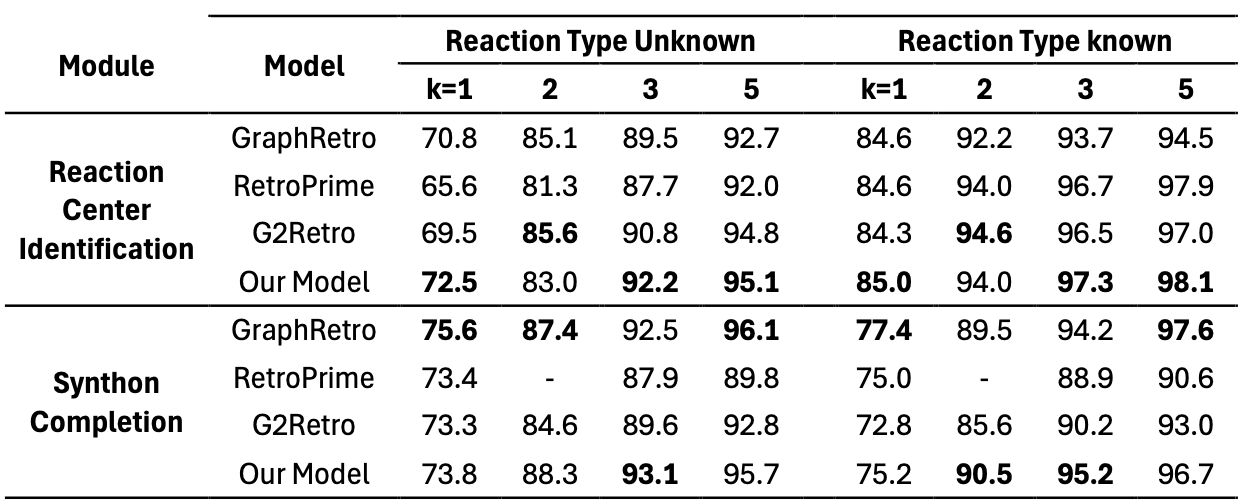} 
 \caption{Top-k module accuracy on USPTO-50K benchmarks}
 \label{tab-3}
\end{table}

\subsection{Interpretation of Reaction Center Identification through Attention Score}
The attention scores generated by the molecular encoder qualitatively reflect the significance of each atom, with reaction centers consistently receiving the highest scores. This demonstrates the model’s ability to leverage molecular representations to identify critical regions of the molecule. As illustrated in \textbf{Fig. 2}, other reactive atoms, even if not directly part of the reaction center, also show elevated scores, indicating the model’s sensitivity to broader chemical contexts.

	In the hydrolysis reaction shown in \textbf{Fig. 2a}, an atom-centered reaction is highlighted where a tertiary butyl group detaches from the oxygen atom in a carboxyl group. The reaction center oxygen atom in the product molecule displays the highest attention score. Additionally, the alpha-carbon and atoms within the carbamide structure exhibit relatively high scores, reflecting their potential as reaction centers and their impact on the model’s attention mechanism. The reductive alkylation reaction in \textbf{Fig. 2b} exemplifies a bond-centered reaction, where a C-N bond is formed. The atom pair constituting the reaction center receives the highest attention scores, demonstrating the model’s ability to focus precisely on the critical bond center. The carbamylation reaction in \textbf{Fig. 2c} illustrates a multi-reaction center scenario. This reaction involves the formation of a carbamide structure through conversion of an isocyanate to a carbamate and amide bond formation. In the product molecule, atoms within the carbamide structure display high attention scores, highlighting their significance in the reaction. After the first step of urea bond cleavage, the model assigns the highest scores to the atoms involved in the changing C-N bond during the intermediate stage. Furthermore, the nitrogen in the primary amine shows elevated scores, indicating its potential involvement despite not being part of the primary reaction center.
    
	These observations underscore the utility of attention score visualization in elucidating the model’s decision-making process. By accurately identifying reaction centers and capturing the broader molecular structure, the model provides interpretable and chemically meaningful predictions, enhancing its reliability and applicability in retrosynthesis tasks.

\begin{figure}[H]
 \centering
 \makebox[\textwidth][c]{\includegraphics[width=1\textwidth]{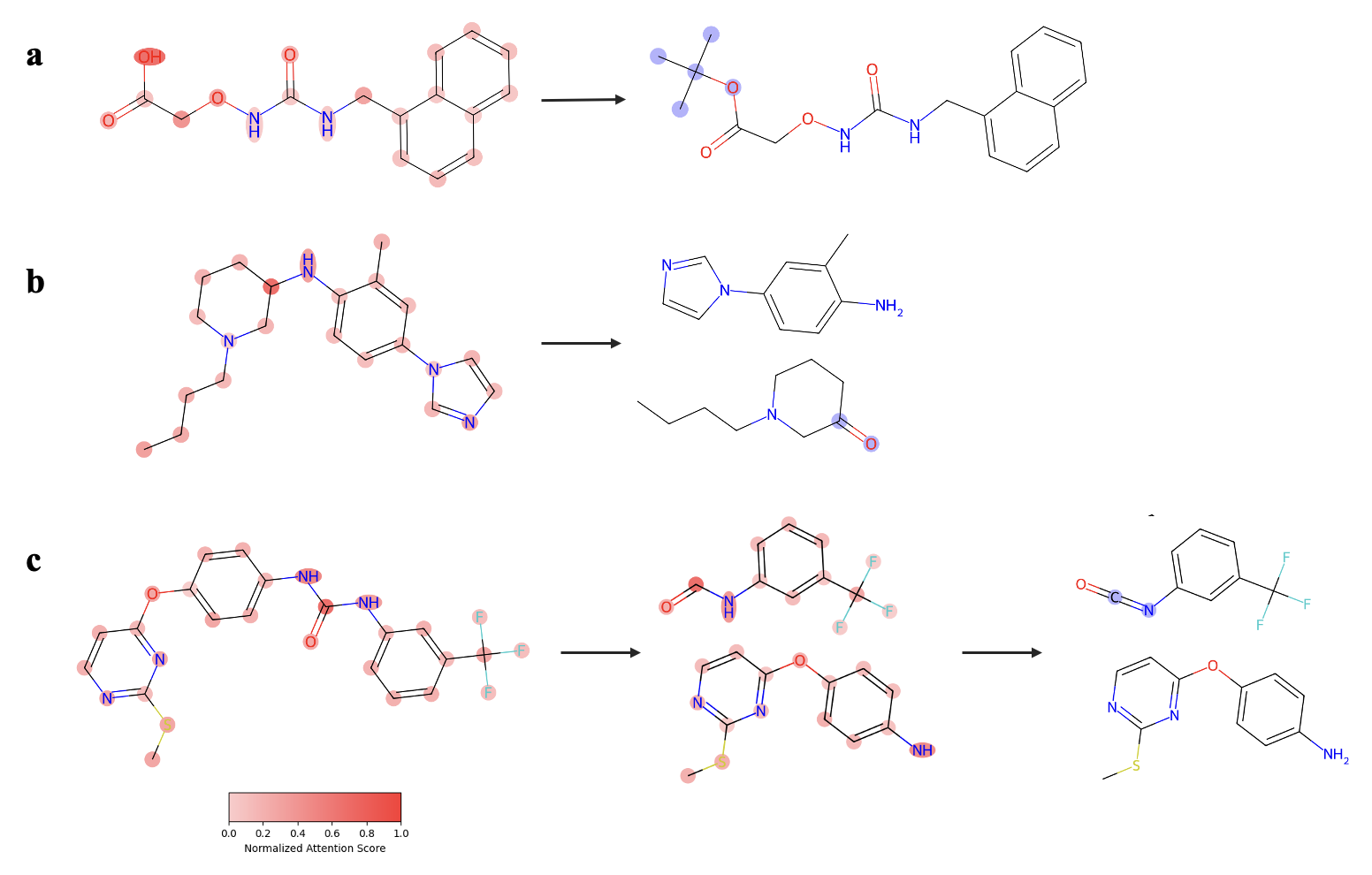}}%
 \caption{\textbf{Interpretation of Reaction Center Identification through Attention Score}, \textbf{a} Hydrolysis reaction with an atom-centered reaction: The oxygen atom in the carboxyl group exhibits the highest attention score, highlighted in red on the product molecule. \textbf{b} Reductive alkylation with a bond-centered reaction: The C-N bond is highlighted, with high attention scores on the atoms at the bond ends. \textbf{c} Carbamylation with multiple reaction centers: Attention scores are distributed across atoms undergoing chemical changes, reflecting the complex nature of the reaction.}
 \label{fig-2}
\end{figure}

\subsection{Hidden Feature Embedding}
The final hidden features extracted from the molecular encoder were visualized using t-distributed Stochastic Neighbor Embedding(t-SNE) to better understand their distribution. \textbf{Fig. 3a} represents the embedding when reaction type information is provided, with points colored according to the reaction type. The visualization shows that the embeddings form well-separated clusters corresponding to the 10 distinct reaction types\cite{b35}, indicating that the model effectively captures the characteristics of each reaction type. Interestingly, some reaction types form separate sub-clusters within the same category. This phenomenon can be attributed to differences in the number of attached hydrogens, which create additional variance within the same reaction type.\cite{b25} 

	\textbf{Fig. 3b} illustrates the distribution of reaction center types, distinguishing between atom centers and bond centers based on their embeddings. The visualization highlights a strong correlation between reaction type and the clustering of reaction centers, as the embeddings align closely with distinct reaction types. This correlation enables the model to leverage reaction type information to more accurately identify reaction centers, contributing to its superior performance in scenarios where reaction type is provided. Notably, 97\% of reactions in the Deprotection cluster are classified as atom centers. Additionally, reaction types such as heteroatom alkylation and arylation, acylation and related processes, C-C bond formation, heterocycle formation, protections, and functional group transformations are exclusively associated with bond centers (100\%). However, reaction types like reduction and oxidation exhibit overlap between atom and bond centers, likely due to a stronger correlation with changes in the number of hydrogen atoms rather than reaction center type. This overlap underscores the complex relationship between reaction features and reaction types in these processes.

\begin{figure}[H]
 \centering
 \makebox[\textwidth][c]{\includegraphics[width=1\textwidth]{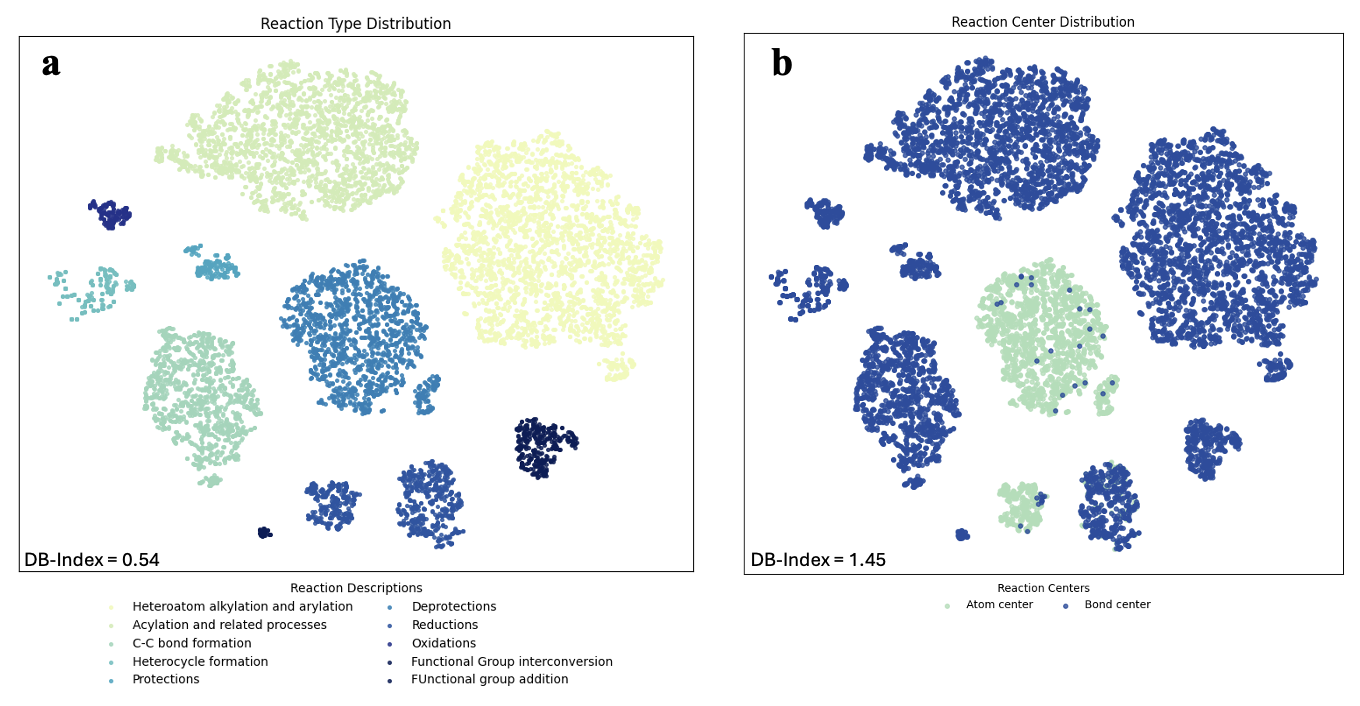}}%
 \caption{\textbf{Hidden Feature Embedding with t-SNE}, \textbf{a} The final hidden features from the molecular encoder trained on USPTO-50k were visualized using t-SNE. The embeddings are colored based on the reaction types from USPTO-50k, showing clear separation across the 10 reaction types. \textbf{b} The same embeddings are visualized based on atom center and bond center types. The distribution strongly correlates with reaction types, with some overlap but generally good separation between the two center types.}
 \label{fig-3}
\end{figure}

\subsection{Multiple-Reaction Center Analysis}
The analysis of multiple reaction centers (RCs) underscores the need for an effective augmentation strategy to mitigate the challenges of limited generalization caused by the USPTO dataset’s constrained size. \textbf{Fig. 4a} compares the distribution of RC counts in the original training set with the augmented trajectory counts. To enrich the dataset and enhance the model’s ability to handle complex reactions, data augmentation was applied by permuting the order of RCs applied in reactions with multiple centers, generating diverse trajectories. To limit excessive augmentation from a single reaction, reactions with four or more RCs were restricted to a maximum of 24 randomly selected trajectories. Reactions with seven or more RCs were excluded due to their heightened complexity, which could impede effective model training.

	\textbf{Fig. 4b} presents the top-5 and top-10 accuracies across reactions with different numbers of RCs. As the number of RCs increases, the length of trajectories generally leads to a decline in accuracy, particularly up to four RCs. However, a notable improvement is observed for reactions with five RCs, likely due to the prevalence of consistent reaction patterns in this category, making them easier for the model to predict accurately.
	While the permutation-based augmentation strategy effectively expanded the dataset and enabled the model to learn a broader variety of RC combinations, challenges remain. The limited number of data points for multiple RCs makes it difficult to evaluate whether the model genuinely generalizes the underlying reaction patterns or simply memorizes them. Nonetheless, the strategy demonstrates significant potential for addressing the challenges associated with multiple RCs, improving the model’s generalization capabilities in such scenarios.

\begin{figure}[H]
 \centering
 \makebox[\textwidth][c]{\includegraphics[width=1\textwidth]{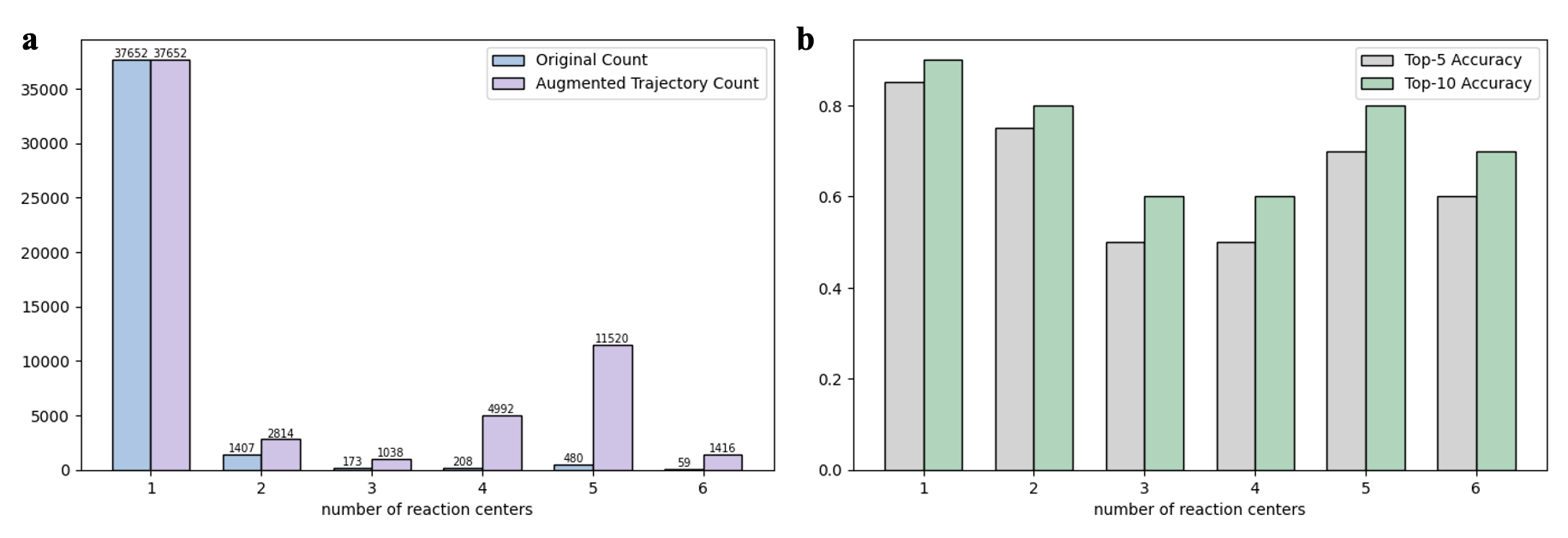}}%
 \caption{\textbf{Multiple-Reaction Center Analysis}, \textbf{a} Number of reactions in USPTO-50k categorized by reaction center count and the corresponding augmented trajectory counts after RC permutation. \textbf{b} Top-5 and top-10 accuracies for reactions with different reaction center counts, illustrating performance trends as the number of reaction centers increases.}
 \label{fig-4}
\end{figure}

\subsection{Multistep Retrosynthesis Predictions}
To further validate the model and assess its practical applications, the model was tested on drug molecules not included in the training dataset. Two FDA-approved drugs from 2023\cite{b44}, Fruquintinib and Nirogacestat, were selected as test cases for evaluation.

	Fruquintinib, a treatment for metastatic colorectal cancer, functions by impeding tumor angiogenesis and growth through the inhibition of VEGFR phosphorylation\cite{b44}. Its synthesis involves an SnAr reaction between 4-chloroquinazoline and 6-hydroxybenzofurane. \textbf{Fig. 5a} illustrates the retrosynthesis approach for Fruquintinib as an example. Our model successfully predicted the reaction pathway, demonstrating its effectiveness in accurately identifying the reaction center and subsequent actions. The process begins with reaction center identification ($P_{RC}=0.5$), successfully pinpointing the C-O bond, which is part of the ether group in Fruquintinib. Subsequently, bond-deletion action at the bond center ($P_{bond}=0.8$) and leaving group Cl attachment at the atom pair of bond center ($P_{atom}=0.4$) are accurately predicted. Finally, the termination step ($P_{term}=0.7$)  is evaluated to determine whether the reaction step concludes. This reaction involved a single reaction center, and the associated actions followed a common reaction pattern, resulting in accurate predictions by the model.
    
Nirogacestat(Ogsiveo), a $\gamma$-secretase inhibitor approved for progressive systemic fibrosis requiring systemic treatment, enhances BCMA-targeted therapies by preventing BCMA cleavage\cite{b44}. Its synthesis involves a series of reductions and amide formation steps. \textbf{Fig. 5b} presents a retrosynthetic prediction for Nirogacestat, where the first pathway successfully mirrored the expected reaction sequence, demonstrating the model’s capability to handle complex multi-step reactions. Our model predicts the reaction centers in Nirogacestat with two potential sites: the C-N bond in the amide group ($P_{RC,1}=0.4$) and the C-N bond in the secondary amine group ($P_{RC,2}=0.1$). The pathway diverges depending on the initial selection of the reaction center, resulting in multiple routes. First reaction center was assigned a higher probability than second one, likely due to the prevalence of amide formation reactions in the training set and the structural similarity around the nitrogen atom in the secondary amine, which distributes probabilities across both C-N bonds. In both pathways, the concurrent occurrence of amide formation and reduction leads to relatively low atom action probabilities. However, the predicted alternate values result in slight variations in leaving group attachments but do not significantly hinder the reaction’s progression. This demonstrates the model’s robustness in handling complex reactions, where minor discrepancies in predictions have minimal impact on the overall retrosynthetic pathway.

\begin{figure}[H]
 \centering
 \makebox[\textwidth][c]{\includegraphics[width=1\textwidth]{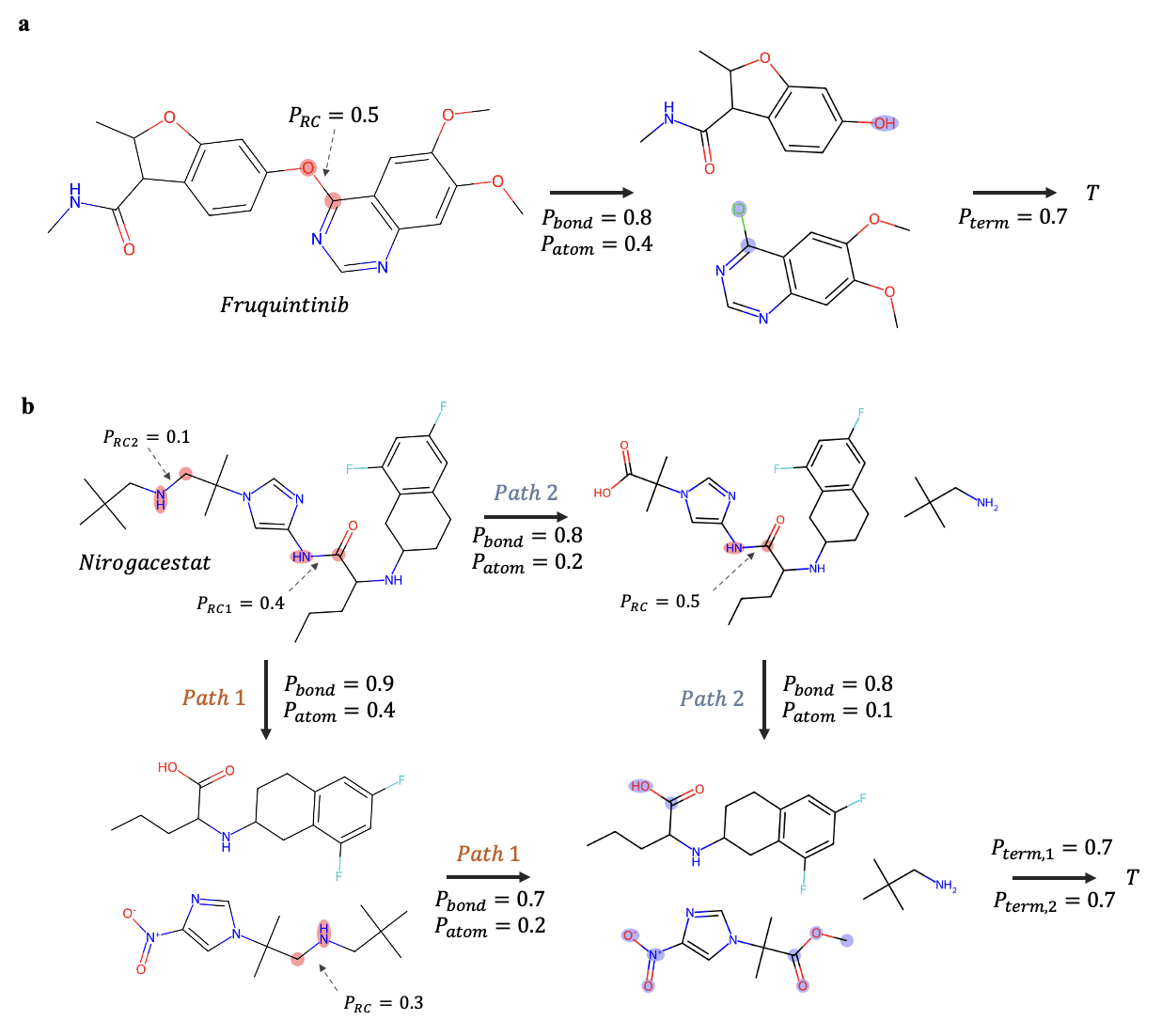}}%
 \caption{\textbf{Retrosynthesis predictions}, \textbf{a} One-step retrosynthesis prediction for Fruquintinib by our model, showing reaction center identification and predicted transformations. \textbf{b} Two-step multi-reaction center retrosynthesis prediction for Nirogacestat.}
 \label{fig-5}
\end{figure}

\section{Discussion}
\subsection{Challenges with Multi-Reaction Center Data}
The limited availability of multi-reaction center data in the USPTO dataset creates a significant obstacle in training robust retrosynthesis models. Multi-reaction center cases often involve intricate dependencies between centers, which are critical for accurately modeling complex reactions. To overcome this, a permutation-based augmentation strategy was introduced, generating diverse trajectories by permuting the order of reaction centers. While this approach enriched the training dataset, it also introduced a limitation: the augmented data included permutations that do not align with actual synthetic pathways. This misalignment can reduce the interpretability of the model’s predictions and may lead to predictions that are synthetically infeasible. To address this, future work could incorporate heuristic or rule-based filters to prioritize permutated trajectories that align with plausible synthetic pathways, ensuring the augmented data better reflects real-world chemistry. These filters could be informed by reaction databases, expert knowledge, or computational predictions of reaction feasibility. By aligning the augmentation process with likely synthesis routes, the model could improve its interpretability and accuracy, reducing the risk of generating synthetically irrelevant predictions.

\subsection{Handling Rare Chemical Actions}
Rare chemical actions, such as uncommon atom group transformations or stereospecific bond changes, pose a significant challenge in retrosynthesis prediction. The uneven distribution of these actions during training often results in underperformance, as the model struggles to learn and generalize from limited examples. To address this, incorporating contextual information for each action, such as atom group properties, could improve the model’s handling of rare cases. By adding action representations, the model can leverage chemical relevance rather than relying solely on data itself, enabling better generalization and more accurate predictions for rare transformations, thus enhancing robustness across diverse and complex chemical reactions.

\section{Conclusion}
In this study, we proposed a hierarchical framework for retrosynthesis prediction, incorporating modules for reaction center identification, action prediction, and termination decision in a systematic manner. Leveraging a molecular encoder pre-trained with contrastive learning and integrating atom- and bond-level information, the model effectively captures general molecular representations. The proposed framework achieved strong performance across key benchmarks, including high top-k accuracies and reaction center identification rates in both reaction type known and unknown scenarios. The framework emphasizes interpretability, with attention score visualizations and embedding distributions showcasing the model’s ability to identify key molecular features and provide chemically meaningful predictions. This explainability builds confidence in the model’s reliability and applicability to retrosynthesis tasks. However, challenges such as the scarcity of multi-reaction center data and handling rare chemical actions persist. Augmentation strategies, like reaction center permutation, were implemented to address data limitations, though further refinement is needed to prioritize realistic pathways. Enriching action representations or adopting task-aware loss functions could mitigate the impact of rare action imbalances. Future work should focus on integrating real-world reaction conditions or reagent-level predictions to bridge the gap between computational and practical retrosynthesis. These developments would enhance the model’s robustness and establish it as a reliable tool for advancing drug discovery and complex molecule design.

\hfill

\textit{Acknowledgments} \\Mention all external funding sources in the acknowledgements.\\

\textit{Disclosure statement:} \\No potential conflict of interest is reported by the authors

\printbibliography
\end{document}